\documentclass[twocolumn,showpacs,preprintnumbers,amsmath,amssymb,epsfig,widetext]{revtex4}
\usepackage{graphicx}% Include figure files
\usepackage{dcolumn}% Align table columns on decimal point
\usepackage{bm}% bold math
\usepackage{epsfig}
\usepackage{color}

\def\fun#1#2{\lower3.6pt\vbox{\baselineskip0pt\lineskip.9pt
  \ialign{$\mathsurround=0pt#1\hfil##\hfil$\crcr#2\crcr\sim\crcr}}}
\def\simgt{\mathrel{\lower0.6ex\hbox{$\buildrel {\textstyle >}
 \over {\scriptstyle \sim}$}}}
\def\simlt{\mathrel{\lower0.6ex\hbox{$\buildrel {\textstyle <}
 \over {\scriptstyle \sim}$}}}

\input epsf

\newcommand{\mnras}{MNRAS}
\newcommand{\apjl}{ApJL}
\newcommand{\aj}{AJL}

\def\be{\begin{equation}}
\def\ee{\end{equation}}
\def\ba{\begin{eqnarray}}
\def\ea{\end{eqnarray}}

\def\nn{\nonumber}

\newcommand{\mpcoh}{\,h^{-1}\,{\rm Mpc}}

\newcommand{\himpc}{{\hbox {$~h^{-1}$}{\rm ~Mpc}}}

\newcommand{\jcap}{J. Cosmology Astropart. Phys.}

\begin{document}

\preprint{}

\title{Broadband Alcock-Paczynski test exploiting redshift distortions}
 
\author{Yong-Seon Song$^{1}$, Teppei Okumura$^{2}$ and Atsushi Taruya$^{3,4}$}
\email{ysong@kasi.re.kr}
\affiliation{$^{1}$ Korea Astronomy and Space Science Institute, Daejeon 305-348, R. Korea \\
$^{2}$ Institute for the Early Universe, Ewha Womans University, Seoul 120-750, R. Korea \\
$^{3}$ Research Center for Early Universe, School of Science, University of Tokyo, Bunkyo-ku, Tokyo 113-0033, Japan \\
$^{4}$ Yukawa institute for theoretical physics, Kyoto University, Kyoto 606-8502, Japan }

\date{\today}

\begin{abstract}
Baryon acoustic oscillations (BAO), known as one of the largest cosmological objects, is now recognized as standard cosmological tool to measure geometric distances via the Alcock-Paczynski effect, by which the observed BAO exhibits characteristic anisotropies in addition to the redshift distortions. This implies that once we know the correct distances to the observed BAO, the tip points of baryon acoustic peaks in the anisotropic correlation function of galaxies, $\xi(\sigma,\pi)$, can form a great circle (hereafter 2D BAO circle) in the $\sigma$ and $\pi$ plane, where $\sigma$ and $\pi$ are the separation of galaxy pair parallel and perpendicular to the line-of-sight, respectively. This 2D BAO circle remains unchanged under the variations of the unknown galaxy bias and/or coherent motion, while it varies transversely and radially with respect to the variations of $D_A$ and $H^{-1}$, respectively. Hereby the ratio between transverse distance $D_A$ and the radial distance $H^{-1}$ reproduces the intrinsic shape of 2D BAO circle, which is {\it a priori} given by the known broadband shape of spectra. All BAO peaks of $\xi(\sigma,\pi)$ are precisely calculated with the improved theoretical model of redshift distortion. We test this broadband Alcock--Paczynski method using BOSS--like mock catalogues. The transverse and radial distances are probed in precision of several percentage fractional errors, and the coherent motion is observed to match with the fiducial values accurately.\end{abstract}

\pacs{draft}

\keywords{Large-scale structure formation}

\maketitle

\section{Introduction}

The observational evidence of ``dark energy" with an effective negative pressure has revolutionized cosmology in the last decade. Since the first evidence of dark energy in 1998 ~\citep{Perlmutter:1998np,Riess:1998cb}, there has been substantial observational and theoretical research aimed at understanding the true nature of this phenomenon. In recent years, many authors have started exploring the possibility that dark energy, and the observed acceleration of the expansion of the Universe, could be the consequence of an incomplete theory of gravity on cosmological scales and may require modifications to Einstein's  theory of General Relativity. A prime goal of precision cosmology in the next decades is to provide cosmological observables in a theoretical model independent way for a fair judgement of confirmation or exclusion. In other words, cosmic observation should be unplugged from our prior knowledge of underlying sciences. We study the theoretical model independent observational tool to probe the distance measures and the growth functions through the Alcock--Paczynski test using the anisotropy spectra of redshift distortions.

The full history of cosmic expansion can be reconstructed using galaxy redshift surveys. Despite the enriched nonlinear structures, the zero-th order description of our current universe is homogeneous and isotropic over sufficiently large scales~\citep{York:2000gk,Peacock:2001gs,Hawkins:2002sg,Percival:2004fs,Zehavi:2004ii,Kazin:2010, Tegmark:2006az,Guzzo:2008ac,Drinkwater:2009sd,Kazin:2009cj,Reid:2009xm,Reid:2012sw}. The measured spatial distribution of galaxies  is determined by the density fluctuations and the coherent peculiar velocities of galaxies. Even though we expect the clustering of galaxies in real space to have no preferred direction, galaxy maps produced by estimating distances from redshifts obtained in spectroscopic surveys reveal an anisotropic galaxy distribution, known as redshift distortions. The anisotropies arise because galaxy recession velocities, from which distances are inferred, include components from both the Hubble flow and peculiar velocities driven by the clustering of matter. Measurements of the anisotropies allow constraints to be placed on the rate of growth of clustering and Hubble flow along the line of sight~\citep{Wang:2006qt,Gaztanaga:2008xz,2008PhRvD..77l3540P,White:2008jy,Song:2010ia,Taruya:2011tz,Samushia:2010ki,Chuang:2011fy,Reid:2012sw,Beutler:2012px,Zhang:2012yt,Anderson:2013vga,Sanchez:2013uxa,Kazin:2013rxa}

The observed spectra in redshift space depend not only on fluctuations of density and velocity fields but also on distance measures of components perpendicular and parallel to the line of sight~\citep{Blake:2003rh,Seo:2003pu}. Unfortunately, those are not simultaneously decomposed out of redshift distortions due to high degeneracy among observables. We can, in principle, resolve this problem if we understand the shape of the power spectrum precisely. In the context of standard cosmology, the shape of spectra is determined before the last scattering surface, and in linear theory, it evolves coherently through all scales. In this case, the shape of spectra is determined by CMB experiments, both the coherent growth functions of density and velocity and the distance measures can be determined separately in precision using the Alcock--Paczynski test in Fourier space~\citep{Song:2012gh}. 

In this paper, we present the Alcock--Paczynski test in the 2D configuration space where galaxy clustering is described by correlation function of $\xi_s(\sigma,\pi)$, where $\sigma$ and $\pi$ are the separation of galaxy pair parallel and perpendicular to the line-of-sight, respectively. If we connect all tip points of BAO peaks observed in $\xi_s(\sigma,\pi)$, it forms a great circle which is called as 2D BAO circle~\citep{2003PhRvD..68f3004H,Matsubara:2004}. The shape of spectra is assumed to be determined at the last scattering surface by CMB experiments. Then the 2D BAO circle becomes invariant to the variation of either density or velocity growth function. However, the 2D BAO circle is altered transversely to the variation of the angular diameter distance $D_A$, and radially to the variation of the inverse of Hubble rate $H^{-1}$. This feature explains us how to measure both transverse and radial distances simultaneously with structure formation. Note that the use of the broadband shape of correlation function or power spectrum in the Alcock--Paczynski test has been already advocated before \citep[e.g.,][]{Ballinger:1996cd,Matsubara:1996nf,Magira:1999bn}, but these previous works have mainly focused on the cases with featureless initial spectrum. Taking account of the BAO as standard ruler, the broadband Alcock--Paczynski test can give a much more impact on the precision estimation of geometric distances. The impact of BAO structure has been studied in many literatures. In configuration space, the cosmological impact of BAO has been mostly explored with the analysis based on the multipole expansion, or angle-averaged quantities such as clustering wedges 
(e.g., \cite{Reid:2012sw}, \cite{Kazin:2012}). 
In this paper, we are particularly interested in the 2D BAO.  
We will discuss the mechanism to break degeneracy among distances and 
growth functions in further detail using the 2D BAO circle. 

There are actually several reasons why we here consider the 2D correlation function. One obvious reason is that the 2D BAO circle in $(\sigma, \pi)$-plane is robust against various systematics, as we mentioned above. This implies that we can in principle measure the quantities, $D_A$ and $H$ in a purely geometrical manner without introducing a detailed model of correlation function. Another reason to use $\xi_s(\sigma,\pi)$, instead of the multipole correlation function, appears when we simultaneously measure the growth of structure with the redshift distortions. With the full 2D correlation function, one can easily distinguish between the geometric and velocity (redshift distortions) effects, clarifying the physical interpretation. In practice, to estimate the growth of structure, a detailed theoretical template of correlation function is necessary, however, the 2D correlation function measured in the binned space of $\sigma$ and $\pi$ is useful to handle non-perturbative systematics such as the Finger of God effect. Just discarding bins which are contaminated by unknown systematics, robust and reliable cosmological analysis is made possible with recently proposed models of redshift distortion based on the perturbation theory calculations \citep{Taruya:2010mx,Reid:2011ar,Vlah:2012}. While the angle-averaged statistics such as those based on the multipole expansion have appeared very popular and been frequently used in the literature, it is generally difficult to exclude the data along the line-of-sight from the angle-averaged quantities. It is thus meaningful to present the analysis of the 2D correlation function, without angular averaging, at least as a complementary method. 

In this paper, based on the perturbation theory model of redshift distortions developed by~\cite{Taruya:2010mx} (see also \citep{Taruya:2013my}), we test our new broadband Alcock--Paczynski test as a tool to measure $D_A$, $H^{-1}$, and the growth function of coherent motion. The improved model used here 
gives a more elaborate description of the correlation function 
than simple factorized formulation. The resultant formula looks very similar 
to the phenomelogical formula frequently used in the literature, but 
includes correction terms at next-to-leading orders, which assist us to 
achieve better fit to the simulated data of dark matter and halo distributions. 
The model further helps to reconstruct the density and velocity power spectra 
simultaneously in an unbiased way \citep{Song:2013}. 
With this improved template, in this paper, 
broadband Alcock--Paczynski test is examined with 
611 BOSS mocking catalogues at $z=0.57$. Those observables are marginalized with unknown galaxy bias and non--perturbative Finger of God effect. The results indicate that the precise measurements of $D_A$, $H^{-1}$, and the growth function of coherent motion are possible by our treatment. Here are layouts of the paper. In section 2, we introduce the improved model of redshift distortions. In section 3, we introduce the mock catalogues which we use. In section 4, we explain the broadband Alcock--Paczynski test. In section 5 and 6, results and conclusion are presented.

\section{The theoretical model of redshift distortions}

\subsection{Needs for improved theoretical model}

As we mentioned in Sec.~1, the redshift--space two--point correlation function of mass tracers, $\xi_s$, exhibits anisotropies mainly due to the peculiar motions of galaxies~\citep{Kaiser:1987qv}, and is described as a function of $\sigma$ and $\pi$, where $\sigma$ and $\pi$ are again the separations between the mass tracers perpendicular and parallel to the line-of-sight direction, respectively. Typically, for the separation along the line of sight (i.e., $\sigma\simeq0$), the clustering in redshift space shows an elongated structure, and the resultant redshift-space correlation function is compressed or supperessed. At small separation, the suppression appears due to the nonlinear velocity dispersions of galaxies, known as Finger of God effect, while the compression in correlation function becomes manifest at large separation, and the clustering pattern is squashed by the coherent infalls toward overdense regions, called Kaiser effect.

In the linear regime of gravitational instability, the density and velocity fields are coherently evolved through the continuity equation. This is exactly the situation characterized by the Kaiser effect, and the redshift-space correlation function $\xi_s(\sigma,\pi)$ is uniquely mapped from the real-space correlation function $\xi(r)$, which is computed with the linear theory of gravitational instability~\citep{1989MNRAS.236..851L,1990MNRAS.242..428M,1991MNRAS.251..128L,1992ApJ...385L...5H,1994MNRAS.266..219F}. Note that the commonly used streaming model~\citep{1995ApJ...448..494F}, in which the function $\xi_s$ is given by the convolution of real-space correlation function with velocity distribution function, also recovers the Kaiser limit of redshift-space correlation in the the large-scale limit. It is, however, important to note that the resultant correlation function $\xi_s(\sigma,\pi)$ differs from the real-space correlation function for all configurations, even those perpendicular to the line of sight~(e.g., \citep{Scoccimarro:2004tg}). This is in marked contrast to the redshift-space power spectrum.

In our previous work~\citep{Song:2013}, as a simple theoretical model, the linear squashing effect is combined in a multiplicative way with the Gaussian damping function characterizing the Finger of God (hereafter FoG) effect. In comparing it with the measured correlation function at $\sigma<60\mpcoh$ and $\pi<40\mpcoh$, the cut--off strategy is applied to eliminate bins in non--linear regime at which the theoretical model of the redshift distortions is broken down. Then the rest of bins at large separations successfully fits to the theory, and provides us coherent motions at low redshifts.

Recently, BOSS released DR9 data, and with this data set, it is now able to uncover the correlation function $\xi_s(\sigma,\pi)$ at larger $\sigma$ and $\pi$, where 2D BAO structure is clearly visible. This implies that making full use of Alcock--Paczynski effect, we may simultaneously measure the geometric distance and the coherent motion field. However, it turned out that measurements of coherent motion from redshift distortion maps suffers from the nonlinear systematic effects. While the BAO basically lies at the scales close to the linear regime, because of the nonlinear gravitational evolution, the small peak structure of BAO tends to be smeared out. Further, the cosmological density and velocity fields couple together, and they evolve nonlinearly. Then, due to its nonlinear nature of the mapping formula between the real and redshift space, the resultant correlation function can get a noticiable nonlinear correction even at BAO scales. Hence, in order to unlock the full cosmological power of the BAO as well as the redshift distortions, a simple factorized model of redshift-space correlation function is insufficient to describe the measured 2D BAO structure, and we need 
a more elaborate description, for which we will discuss below.

\subsection{The improved theoretical spectra in Fourier space}

Before presenting an improved model of correlation function, let us first consider the redshift-space power spectrum. On large scales, the observed large-scale structure is basically described by a small perturbation to the homogeneous universe. Ignoring the higher-order contributions, the two-dimensional power spectrum in redshift space, $\tilde{P}(k,\mu)$, given as the function of wavenumber $k$ and directional cosine $\mu$ between line-of-sight direction and $k$, is solely characterized by the Kaiser effect, and can be expressed as~\cite{Kaiser:1987qv}
\begin{equation}
\tilde{P}(k,\mu) = P_{\delta\delta}^{\rm lin}(k) + 2\mu^2 P_{\delta\Theta}^{\rm lin}(k) + \mu^4 P_{\Theta\Theta}^{\rm lin}(k).
\label{eq:Kaiser}
\end{equation}
As we mentioned, the observed redshift-space power spectra suffers significantly from nonlinear corrections even at large scales. One important effect is the gravitational evolution. Further, the random motion of galaxies is known to cause the FoG effect. The approximate prescription to describe these nonlinear effects was proposed by Scoccimarro, and it is expressed as~\citep{Scoccimarro:2004tg}
\ba
\tilde{P}(k,\mu) = \left\{P_{\delta\delta}(k) + 2\mu^2 P_{\delta\Theta}(k) + \mu^4 P_{\Theta\Theta}(k)
\right\}G^{\rm FoG}(k\mu\sigma_p).
\label{eq:S04}
\ea
where $\sigma_{\rm p}$ is the one-dimensional velocity dispersion. Nonlinear corrections due to the random motion is described by the factor $G^{\rm FoG}$ which is given in the Gaussian function,
\ba
G^{\rm FoG}(k\mu\sigma_p) = \exp\left\{-(k\mu\sigma_{\rm p})^2\right\},
\label{eq:Ggauss}
\ea
The auto- and cross-power spectra of the two fields expressed as $P_{XY}^{\rm lin}(k)$, with $X$ and $Y$ being either $\delta$ or $\Theta$, are the nonlinear generalization of the linear counterpart $P_{XY}^{\rm lin}(k)$ in Eq.~\ref{eq:Kaiser}. With the model given by Eq.~\ref{eq:S04}, the prediction of redshift-space correlation function $\xi_s(\sigma,\mu)$ reproduces reasonalby well the anisotropic clustering structure at scales of $\sigma<60 \mpcoh$ and $\mu<30\mpcoh$. However, it fails to match the observed 2D BAO peak structure. 

Beyond a simple model prescription, one crucial aspect of the redshift distortions may be that the linear squeezing and nonlinear smearing effects on distorted maps are not separately treated. Taking account of this fact, Taruya, Nishimichi \& Saito~\citep{Taruya:2010mx} proposed an improved model of the redshift-space power spectrum, in which the coupling between the density and velocity fields associated with the Kaiser and the FoG effects is perturbatively incorporated into the power spectrum expression. The resultant expression looks similar to Eq.~\ref{eq:S04}, but includes nonlinear corrections consisting of 
higher-order polynomials~\citep{Taruya:2010mx}:
\ba
\tilde{P}(k,\mu) &=& \big\{P_{\delta\delta}(k) + 2\mu^2 P_{\delta\Theta}(k) + \mu^4 P_{\Theta\Theta}(k) \nn\\
&+& A(k,\mu) + B(k,\mu)\big\}G^{\rm FoG}.
\label{eq:TNS10}
\ea
Here the $A(k,\mu)$ and $B(k,\mu)$ terms are the nonlinear corrections, and are expanded as power series of $\mu$, including the powers up to $\mu^6$ for the $A$ term and $\mu^8$ for the $B$ term. In this paper, adopting the model given in Eq.~\ref{eq:TNS10}, we will proceed to the cosmological analysis, and investigate the impact of these nonlinear corrections.

\begin{figure}
\begin{center}
\resizebox{3.3in}{!}{\includegraphics{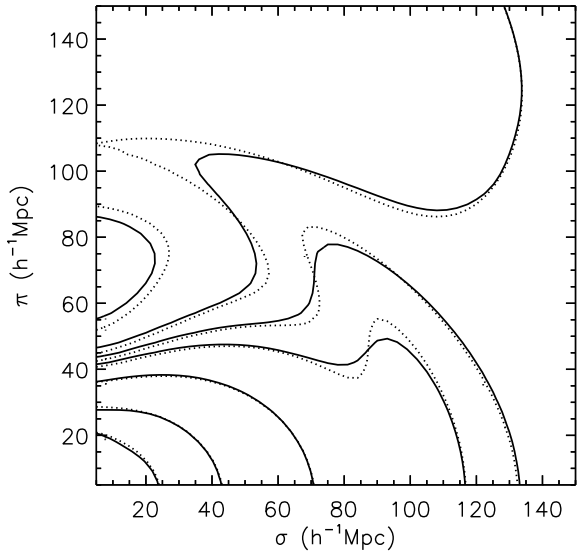}}
\end{center}
\caption{Dotted contours represent the theoretical $\xi_s(\sigma,\pi)$ of the simple combination of Kaiser model and Gaussian FoG effect. Solid contours represent the improved $\xi_s(\sigma,\pi)$. The level of $\xi_s(\sigma,\pi)$ is (0.2, 0.06, 0.16, 0.005, 0.002, -0.001, -0.006) from the inner to outer contours.}
\label{fig:corrections}
\end{figure}

In computing the power spectrum, we need to properly take into account the effect of nonlinear gravitational evolution not only for the $A$ and $B$ terms but also for the auto- and cross-power spectra $P_{XY}(k)$. Since the standard perturbation theory is known to produce ill-behaved epansion leading to the bad UV behavior, a consistent calculation of correlation function should be made with the improved perturbation theory with an appropriate UV regularization. Here, we apply the resummed perturbation theory called {\tt RegPT}~\citep{Taruya:2012ut}, and following the prescription described in~\cite{Taruya:2013my}, we compute the power spectra $P_{XY}(k)$ as well as the $A$ and $B$ terms, including the nonlinear corrections up to the two-loop order, i.e., next-to-next-leading order. Note that the {\tt RegPT} shceme is based on a multipoint propagator expansion, and with this scheme, any statistical quantities consisting of density and velocity fields are built up with multipoint propagators, in which nonperturbative properties of gravitational growth  are wholly encapsulated~\citep{Bernardeau:2008fa}. Making use of the analytic properties of the propagators, a novel regularized treatment has been implemented~\citep{Bernardeau:2011dp}, showing that the proposed scheme can be used to give a percent-level prediction of power spectrum and correlation function at weakly nonlinear regime in both real and redshift spaces~\citep{Taruya:2012ut,Taruya:2013my}.

With the {\tt RegPT} scheme, we pre-compute the nonlinear corrections to the auto- and cross-power spectra $P_{XY}$ in the fiducial cosmological model:
\ba\label{eq:closure}
\delta P_{XY}^{\rm fid}(k) = P_{XY}^{\rm fid}(k) - P_{XY}^{\rm fid, lin}(k).
\ea
Adding up linear and nonlinear parts, we have
\ba
P_{XY}(k) = \frac{G_X(z)}{G_X^{\rm fid}(z)}\frac{G_Y(z)}{G_Y^{\rm fid}(z)}\left\{P_{XY}^{\rm fid, lin}(k)+\delta P_{XY}^{\rm fid}(k)\right\}.
\ea
Here, the prefactor $G_X(z)$ includes the information on the structure growth and galaxy bias, which will be later clarified in Sec.~IV-A. Strictly speaking, the second term depends on not only the linear power spectrum $\delta_{XY}^{\rm fid}$ but also the cosmological model as well as the gravity theory. %At the weakly nonlinear scales of our interest, however, the term $\delta P_{XY}$ itself is a small contribution to the total power spectrum, and a slight deviation from fiducial model would not change drastically the nonlinear corrections. We simply let this term unchanged from its fiducial value.  

To sum up, we adopt the redshift-space power spectrum, $\tilde{P}(k,\mu)$, given in Eq.~\ref{eq:TNS10}, which can be recast as:
\ba
\label{eq:pkred_in_Q}
\tilde{P}(k,\mu) =\sum_{n=0}^8\,Q_{2n}(k)\mu^{2n}\,G^{\rm FoG}(k\mu\sigma_p)\,,
\ea
Here, $\sigma_p$ is the free parameter which we allow to float in the cosmological analysis. Although this may lead to an additional uncertainty in the parameter estimation, our previous analysis suggests that as long as we consider the weakly nonlinear scales, cosmological analysis can be made independently of the functional form of FoG effect. The functions $Q_{2n}$ are given by
\ba\label{eq:Q}
Q_0(k)&=&P_{\delta\delta}^{\rm lin}(k) +\delta P_{\delta\delta}(k),\nn\\
Q_2(k)&=&2P_{\delta\Theta}^{\rm lin}(k) +2\delta P_{\delta\Theta}(k)+ C_2(k),\nn\\
Q_4(k)&=&P_{\Theta\Theta}^{\rm lin}(k)+\delta P_{\Theta\Theta}(k)+ C_4(k),\nn\\
Q_6(k)&=&C_6(k),\nn\\
Q_8(k)&=&C_8(k),
\ea
where $C_n$ includes the nonlinear correction terms $A$ and $B$, and we assume the perfect correlation between linear density and velocity fields, i.e., $P_{\delta\Theta}^{\rm lin}(k)=[P_{\delta\delta}^{\rm lin}(k)P_{\Theta\Theta}^{\rm lin}(k)]^{1/2}$. Apart from linear-order quantities, nonlinear contributions in each term are computed with the {\tt RegPT} scheme, consistently including the perurbative corrections up to the two-loop order. 

\begin{figure}
\begin{center}
\resizebox{3.3in}{!}{\includegraphics{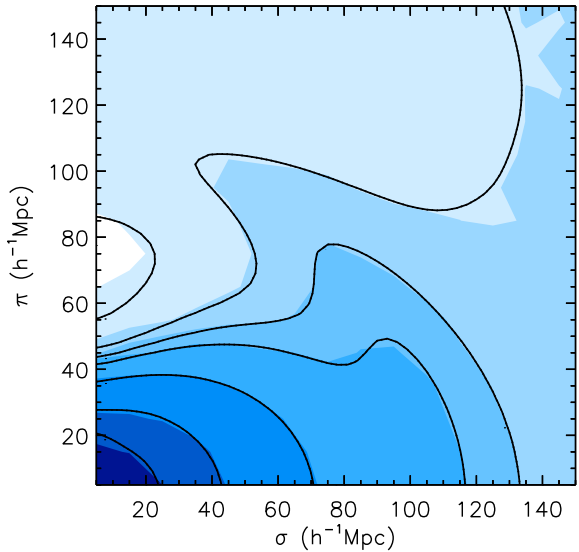}}
\end{center}
\caption{
We compare the observed and the theoretical $\xi_s(\sigma,\pi)$. The dotted red contours represent the observed $\xi_s(\sigma,\pi)$ using simulated maps. The black contour in the left panel represents the theoretical $\xi_s(\sigma,\pi)$ without using $\delta P_{XY}$ and $C_n$, and the black contour in the right panel represents the theoretical $\xi_s(\sigma,\pi)$ using theoretical models described in Section 2.}
\label{fig:xi_all}
\end{figure}

\subsection{The improved correlation $\xi_s(\sigma,\pi)$ in configuration space}

Once provided the power spectrum, it is now rather easy task to compute the correlation function. The redshift-space correlation function $\xi_s(\sigma,\mu)$ is generally expanded as
\ba\label{eq:xi_eq}
\xi_s(\sigma,\pi)&=&\int \frac{d^3k}{(2\pi)^3} \tilde{P}(k,\mu)e^{i{\bf k}\cdot{\bf s}}\nn\\
&=&\sum_{\ell:{\rm even}}\xi_\ell(s) {\cal P}_\ell(\nu)\,,
\ea
with ${\cal P}$ being the Legendre polynomials. Here, we define $\nu=\pi/s$ and $s=(\sigma^2+\pi^2)^{1/2}$. The moments of correlation function, $\xi_\ell(s)$, are defined by,
\ba
\xi_\ell(s)=i^\ell\int\frac{k^2dk}{2\pi^2}\,\tilde{P}_\ell(k)\,j_\ell(ks)\,.
\ea
For the improved model given in Eq.~\ref{eq:pkred_in_Q}, the multipole power spectra $\tilde{P}_\ell(k)$ are explicitly given by,
\ba
\tilde P_0(k)&=&p_0(k),\nn\\
\tilde P_2(k)&=&\frac{5}{2}\left[3p_1(k)-p_0(k) \right],\nn\\
\tilde P_4(k)&=&\frac{9}{8}\left[35p_2(k)-30p_1(k)+3p_0(k)\right],\nn\\
\tilde P_6(k)&=&\frac{13}{16}\left[231p_3(k)-315p_2(k)+105p_1(k)-5p_0(k)\right],\nn\\
\tilde P_8(k)&=&\frac{1}{64}\left[6435p_4(k)-12012p_3(k)+6930p_2(k)\right.
\\
&&\qquad\qquad\qquad\qquad\qquad\left.-1260p_1(k)+35p_0(k)\right]\nn,
\ea
where we define the function $p_m(k)$: 
\ba
p_m(k)&=&\frac{1}{2}\sum_{n=0}^4\frac{\gamma(m+n+1/2,\kappa)}{\kappa^{m+n+1/2}}\,Q_{2n}(k)
\ea
with $\kappa=k^2\sigma_p^2$. The function $\gamma$ is the incomplete gamma function of the first kind: 
\ba
\gamma(n,\kappa)=\int^{\kappa}_0dt\, t^{n-1}\,e^{-t}\,.
\ea

In Fig.~\ref{fig:corrections}, we plot the redshift-space correlation function in $(\sigma,\pi)$-plane, and compare the improved model based on the expression \ref{eq:TNS10} (or Eq.~\ref{eq:pkred_in_Q}) with simple Scoccimarro's model given in Eq.~\ref{eq:S04}. Dashed contours represent the result of the simple combination of Kaiser and Gaussian FoG effects, while solid contours represent the improved model prediction. Here, in both cases, the constant bias parameter and the velocity dispersion parameter $\sigma_p$ are set to the best-fitting values obtained from the cosmological analysis with Mock catalog (see Table 1). At the scales accessible to the previous SDSSII, i.e., $\sigma<60\mpcoh$ and $\pi<40\mpcoh$, both theoretical models agree to each other beyond the cut--off scales. At larger scales, $\sigma<150\mpcoh$ and $\pi<150\mpcoh$, however, there appears noticiable discrepancy between both descriptions. Differences are particularly manifest around the baryon acoustic peak. Note that in computing the improved model, the contribution from the $Q_8$ term turns out to be small in correlation function, and no serious difference appears if we neglect such a higher-order polynomial. Hence, in the analysis below, we drop the $Q_8$ term  and compute the correlation function. As our theoretical model is limited at large scales, the cut--off scales are applied at $s=50\mpcoh$ and $\pi=20\mpcoh$.

\section{Measurements}

\begin{figure}
\begin{center}
\resizebox{3.2in}{!}{\includegraphics{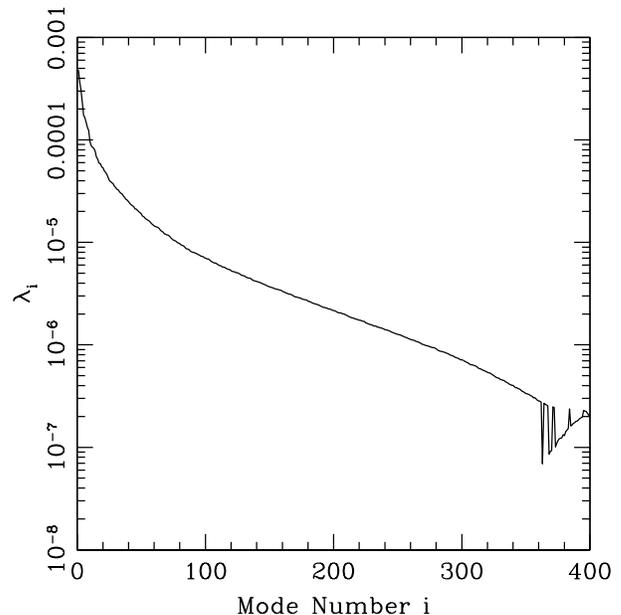}}
\end{center}
\caption{Eigenvalues for the covariance matrix for the measurement of the correlation function of the CMASS galaxy mock catalogs.}
\label{fig:eigenmode}
\end{figure}

\subsection{Measured correlation function from mock catalogs}

In order to check the validity of our improved model in Sec.~2, we first present the redshift-space correlation function $\xi_s(\sigma,\pi)$ measured from a mock galaxy sample. We use the mock galaxy catalogs created by \cite{Manera:2013}, which is designed to investigate the various systematics in a galaxy sample from Data Release 9 (DR9) of the Baryon Oscillation Spectroscopic Survey (BOSS) \citep{Schlegel:2009,Eisenstein:2011,Anderson:2012}, referred to as ``CMASS" galaxy sample. In constructing the mock galaxy catalog, \cite{Manera:2013} utilized second-order Lagrangian perturbation theory (2LPT) for the galaxy clustering driven by gravity, and thus 
it enables to create a mock catalog much faster than running an $N$-body simulation. The redshift range of galaxies in the catalog is $0.43<z<0.7$ and each catalog contains $\sim 2.67\times 10^5$ galaxies, 90\% of which are central galaxies residing in dark matter halos of $\sim 10^{13}h^{-1}M_\odot$. We analyze $611$ realizations of those mock CMASS catalogs. 

We measure the two-dimensional redshift-space correlation function using the Landy-Szalay estimator \citep{Landy:1993}: 
\begin{equation}
\xi_s(\sigma,\pi)=\frac{DD-2DR+RR}{RR},
\end{equation}
where $DD$, $RR$, and $DR$ are the normalized counts of galaxy-galaxy, random-random, and galaxy-random
pairs, respectively, in a particular bin in the space of $(\sigma,\pi)$. 
The random catalog, which was also created by \cite{Manera:2013}, contains $1.76\times 10^6$ points as galaxies with no influence of gravitational clustering. 

The resulting redshift-space correlation function measured from the mock catalogs, after averaging over all realizations, is shown in Fig.~\ref{fig:xi_all}. The filled blue contours are the measured correlation function, which is pixelized into rectangular cells with $\Delta\sigma = \Delta \pi = 10\himpc$. The contour is essentially the same as figure 3 of \cite{Reid:2012sw}, where they used the observed data of CMASS galaxies. Note that the ridge structure seen at scales around $\sim 105\himpc$ is the two-dimensional feature of baryon acoustic oscillations (BAO) \citep{Matsubara:2004}, which has been already detected for several observational data \citep{Okumura:2008, Gaztanaga:2008xz, Kazin:2010, Reid:2012sw}. 

In Fig.~\ref{fig:xi_all}, we also show the improved model prediction of $\xi_s(\sigma,\pi)$, depicted as solid contours. The improved model successfully describes the ridge structure, which is somewhat smeared out due to the nonlinear effects of gravity and redshift distortions. This fact implies that with the improved model of correlation function, we can handle the broad-band shape of the correlation function. It enables us to put a tigher constraint on the geometric distances through the Alcock-Paczynski effect, which we will discuss in next section.

\subsection{Covariance matrix}
Because there are strong correlations between different bins of the correlation function $\xi_s(\sigma,\pi)$, 
it is necessary to estimate a covariance matrix. 
Jackknife resampling \citep{Lupton:1993}, which is straightforward to perform and thus is often adopted, 
tends to underestimate cosmic variance at BAO scales we are interested in. 
Since we want to analyze the full two-dimensional map that has much larger degree-of-freedom, 
we need hundreds of independent realizations in order to have the non-singular covariance matrix. 
In our analysis we use the mock CMASS catalogs directly in order to estimate the covariance matrix because 
the total number of the catalogs, 611, is large enough. 
In fact the mock catalogs were created for the purpose of estimating the covariance. 
However, in previous works different statistics which have smaller degree of freedom were used, 
e.g., multipoles \citep{Reid:2012sw}.

Using the correlation function computed from each of the 611 mock catalogs, we estimate a covariance matrix as 
\begin{equation}
{\rm Cov}(\xi_i,\xi_j)=\frac{1}{N-1}\sum^{N}_{k=1}[\xi_k({\bf r}_i)-\overline{\xi}({\bf r}_i)][\xi_k({\bf r}_j)-\overline{\xi}({\bf r}_j)],
\end{equation}
where $N=611$, $\xi_k({\bf r}_i)$ represents the value of the correlation function of $i$th bin of $(\sigma,\pi)$ in $k$th 
realization, and $\overline{\xi}({\bf r}_i)$ is the mean value of $\xi_k({\bf r}\_i)$ over realizations. 
We can then obtain the correlation matrix as 
\begin{equation}
C_{ij}=\frac{{\rm Cov}(\xi_i,\xi_j)}{\sqrt{{\rm Cov}(\xi_i,\xi_i){\rm Cov}(\xi_j,\xi_j)}}.
\end{equation}
In order to reduce the statistical noise in our covariance matrix, we perform a singular value decomposition (SVD)
of the matrix as done in \cite{Song:2010, Song:2011},
\begin{equation}
 C_{ij} =  U_{ik}^{\dag} D_{kl} V_{lj},
 \end{equation}
 where $U$ and $V$ are orthogonal matrices that span the range and
the null space of $C_{ij}$ and $D_{kl}=\lambda^2\delta_{kl}$, a diagonal
 matrix with singular values along the diagonal. The details of SVD method is explained in Appendix A.

In Fig.~\ref{fig:eigenmode}, we show the eigenvalues ($\lambda_i$) for
the increasing eigenmodes. Since higher-order modes are dominated by noise, we will truncate 
such eigenmodes as seen in next section. 

\begin{figure*}
\begin{center}
\resizebox{3.2in}{!}{\includegraphics{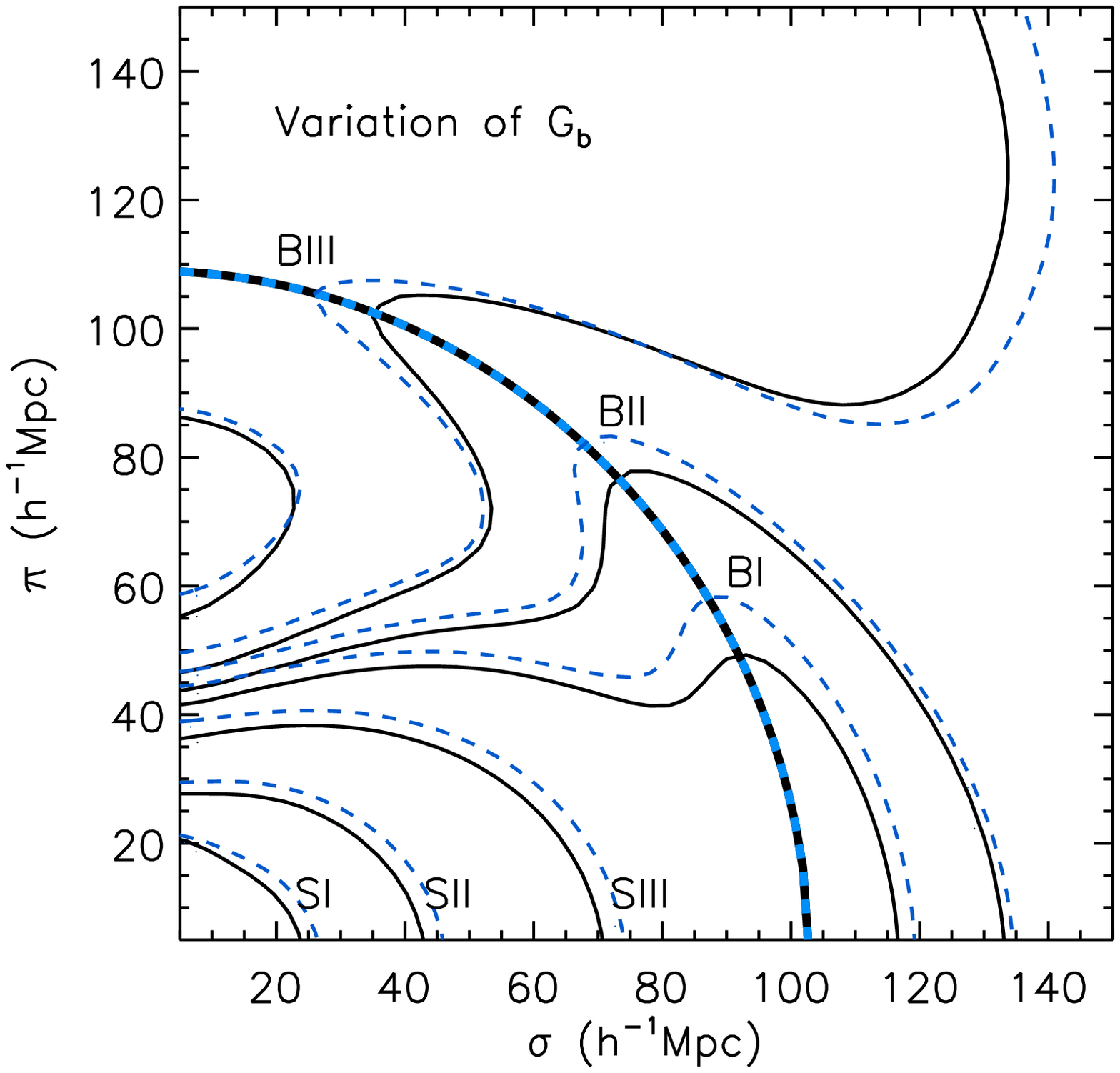}}\hfill
\resizebox{3.2in}{!}{\includegraphics{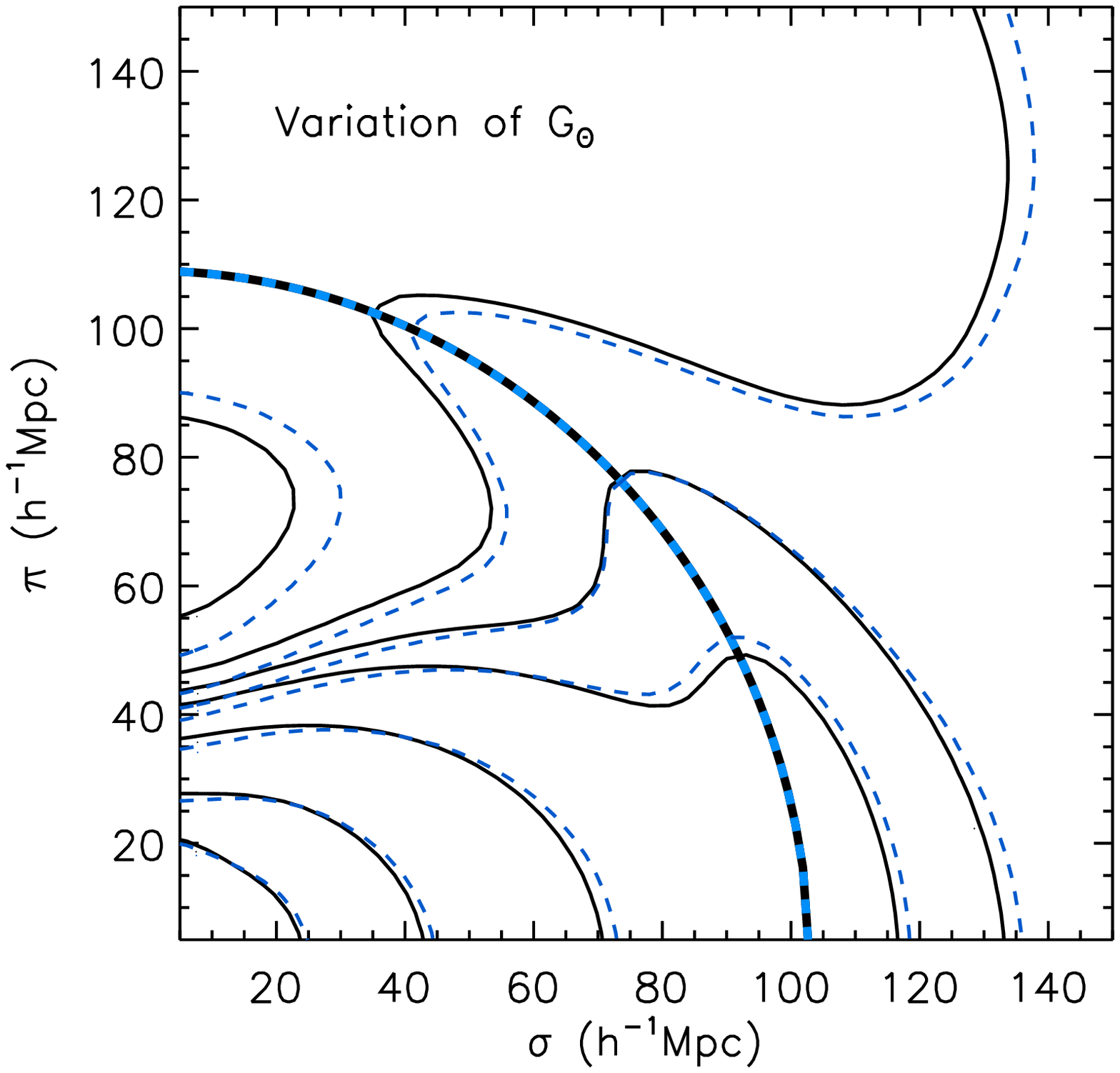}}
\end{center}
\caption{{\it (Left panel)} Thin black solid curves represent the contours of fiducial $\xi_s(\sigma,\pi)$ with the same levels in~Fig.\ref{fig:corrections}. Three inner curves labeled by SI, SII and SIII represent the contours without BAO peaks, and three outer curves labeled by BI, BII and BIII represent the contour with BAO peaks. Thick black solid curve represents the fiducial 2D BAO circle on which all BAO peaks are located. Thin blue dash curves represent the contours of $\xi_s(\sigma,\pi)$, when we vary $G_b$ by 10$\%$. Thick blue dash curve represents the 2D BAO circle with variation of $G_b$. {\it (Right panel)} Thin black solid curves represent the contours of fiducial $\xi_s(\sigma,\pi)$ with the same levels in~Fig.\ref{fig:corrections}. Thick black solid curve represents the fiducial 2D BAO circle on which all BAO peaks are located. Thin blue dash curves represent the contours of $\xi_s(\sigma,\pi)$, when we vary $G_{\Theta}$ by 10$\%$. Thick blue dash curve represents the 2D BAO circle with variation of $G_{\Theta}$.}
\label{fig:con_growth}
\end{figure*}

\section{Statistical method to test Alcock--Paczynski effect using 2D BAO}

\subsection{Fitting parameters}
The observed clustering in redshift space is known as a probe of structure formation of density and velocity fields. In addition, it can be a tracer to determine both transverse and radial distances exploiting Alock--Paczynski effect~\citep{Blake:2003rh,Seo:2003pu,Wang:2006qt,Song:2011,Chuang:2011fy,Beutler:2012px}. Unfortunately, there is significant degeneracy among unknowns. None of observables are measured in precision, unless the specific condition is imposed {\it a priori}. Couple of approaches are proposed to break this degeneracy in~\cite{Song:2012gh}; one is the treatment to give a broadband prior on the shape of spectra determined by CMB experiments, and the other is the treatment to unify the transverse distance with the radial distance based upon the assumption of  FLRW universe prior. Here, we investigate the approach with the given broadband shape prior.

In the context of standard cosmology, the shape of spectra is determined before the last scattering surface, and in linear theory, it evolves coherently through all scales. The history of structure formation evolution is divided into two regimes; epochs before matter-radiation equality ($a_{eq}$) and a later epoch of coherent evolution of unknown effect on structure formation from new physics. We can then express various power spectra of the density field splits into these two epochs, with the shape-dependent part determined by knowledge of our standard cosmology, and the coherent evolution part only affected by new physics. Mathematically, this is written as, 
\ba
P_{\Phi\Phi}(k,a)&=&D_{\Phi}(k)g_{\Phi}^2(a),\nn\\
P_{bb}(k,a)&=&D_{m}(k)g_{b}^2(a),\nn\\
P_{\Theta_m\Theta_m}(k,a)&=&D_m(k)g_{\Theta_m}^2(a)\,,
\ea
where $\Phi$ denotes the curvature perturbation in the Newtonian gauge, 
\begin{equation}
 ds^2=-(1+2\Psi)dt^2+a^2(1+2\Phi)dx^2\,.
\end{equation}
These power spectra are then partitioned into a scale--dependent part ($D_{\Phi}(k),D_{m}(k)$) and a scale-independent (coherent evolution) component ($g_{\Phi}$, $g_{b}$, $g_{\Theta_m}$). We define here $g_{b}=b\,g_{\delta_m}$ where $b$ is the standard linear bias parameter between galaxy (or cluster) tracers and the underlying dark matter density. The expression of $D_{\Phi}(k)$ and $D_m(k)$ is available in citation, and assumed to be given by CMB experiments precisely. The $D_m(k)$ is determined by the fiducial cosmological parameters.

\begin{figure*}
\begin{center}
\resizebox{3.2in}{!}{\includegraphics{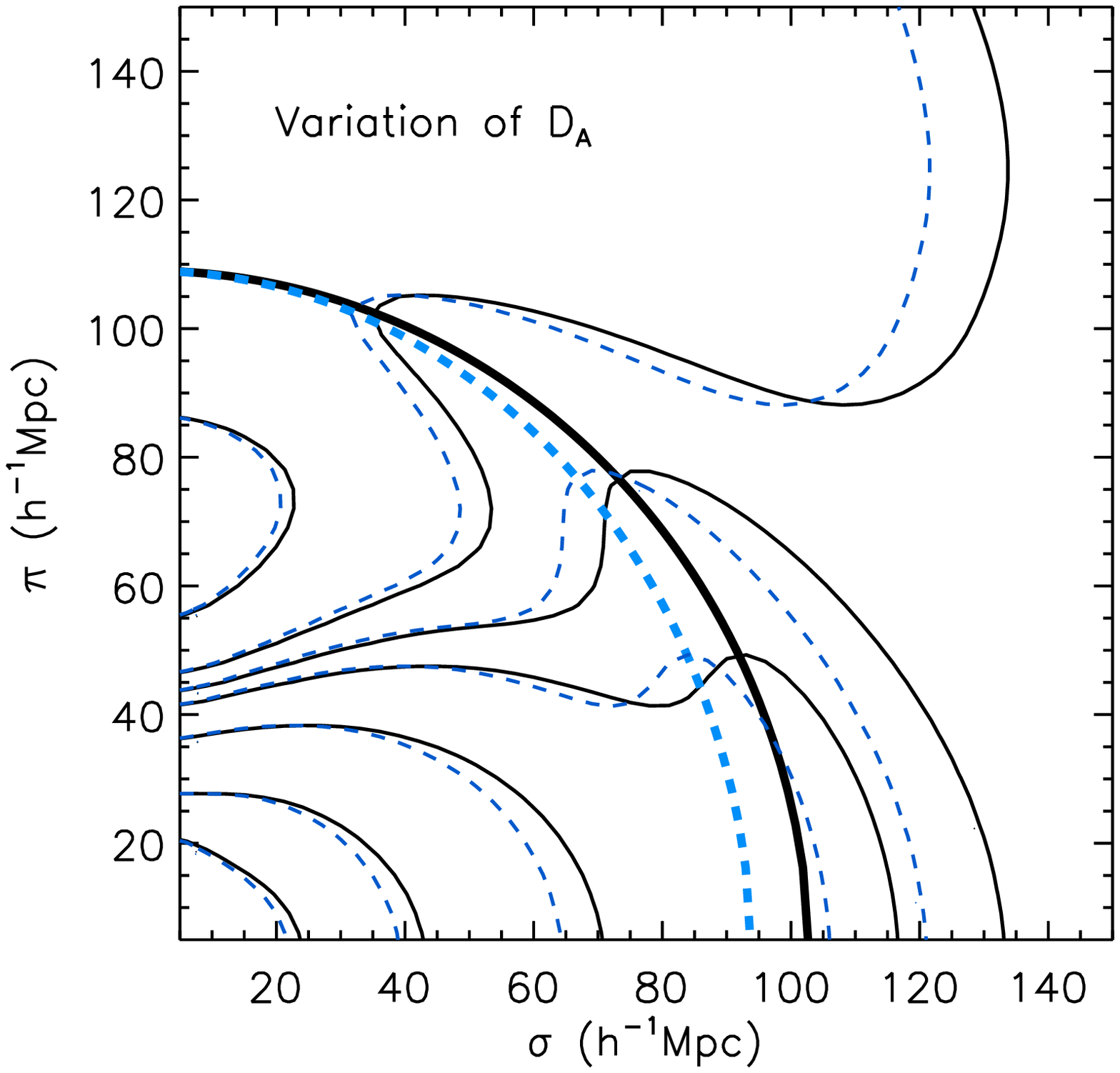}}\hfill
\resizebox{3.2in}{!}{\includegraphics{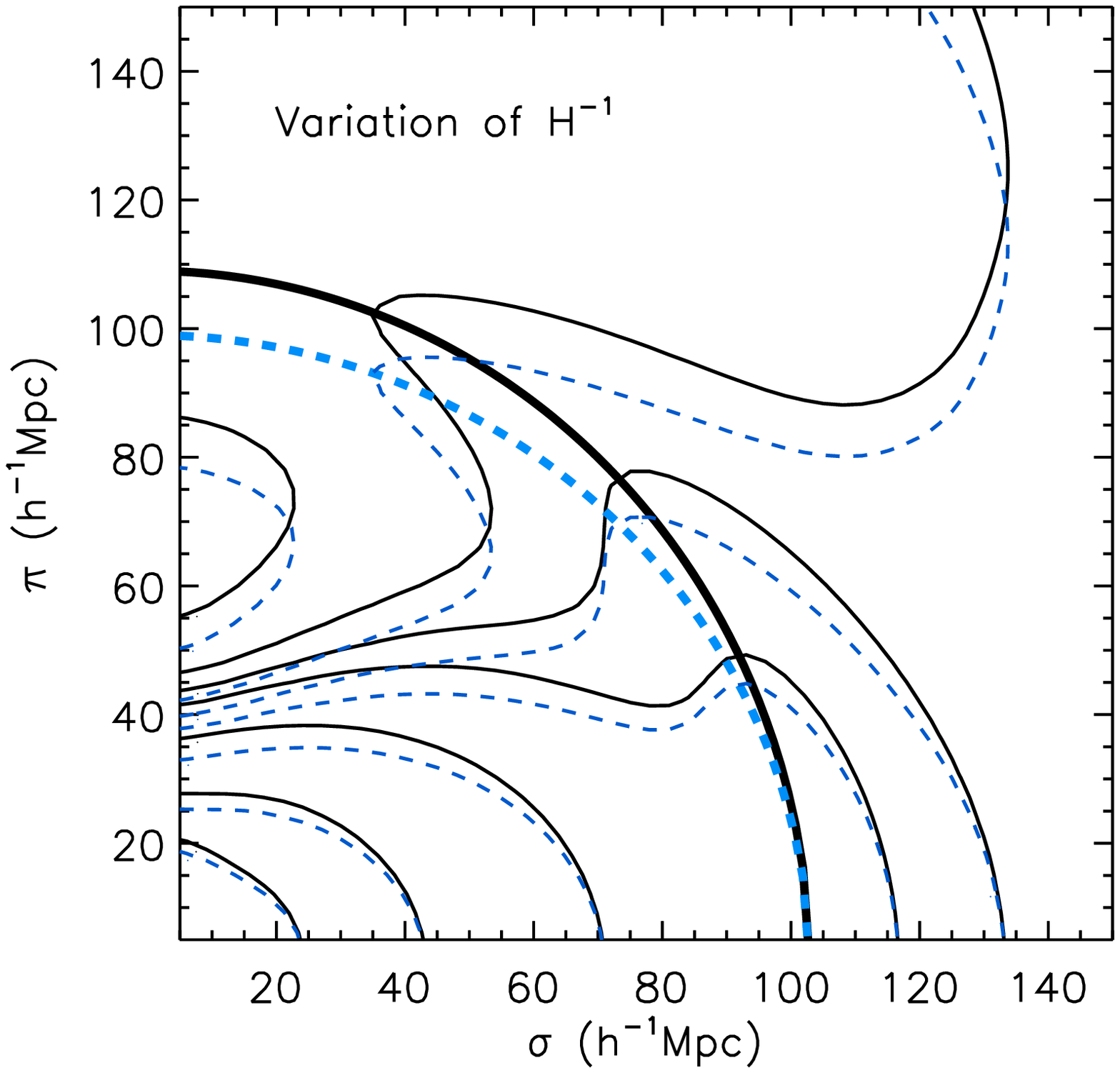}}
\end{center}
\caption{{\it (Left panel)} Thin black solid curves represent the contours of fiducial $\xi_s(\sigma,\pi)$ with the same levels in~Fig.\ref{fig:corrections}. Thick black solid curve represents the fiducial 2D BAO circle on which all BAO peaks are located. Thin blue dash curves represent the contours of $\xi_s(\sigma,\pi)$, when we vary $D_A$ by 10$\%$. Thick blue dash curve represents the 2D BAO circle with variation of $D_A$. {\it (Right panel)} Thin black solid curves represent the contours of fiducial $\xi_s(\sigma,\pi)$ with the same levels in~Fig.\ref{fig:corrections}. Thick black solid curve represents the fiducial 2D BAO circle on which all BAO peaks are located. Thin blue dash curves represent the contours of $\xi_s(\sigma,\pi)$, when we vary $H^{-1}$ by 10$\%$. Thick blue dash curve represents the 2D BAO circle with variation of $H^{-1}$.}
\label{fig:con_distance}
\end{figure*}

Unlike the shape part, the coherent evolution component, $g_{\Phi}$, $g_{b}$ and $g_{\Theta_m}$ are not generally parameterized by known standard cosmological parameters. We thus normalize these growth factors at $a_{eq}$ such that,
\ba
g_{\Phi}(a_{eq})&=&1,\nn\\
g_{\delta_m}(a_{eq})&=&a_{eq}g_{\Phi}(a_{eq}), \nn\\
g_{\Theta_m}(a_{eq})&=&-\frac{dg_{\delta_m}(a_{eq})}{d\ln a}\,.
\ea
Instead of determining growth factors using cosmological parameters, we measure these directly in a model-independent way at the given redshift without referencing to any specific cosmic acceleration model and with the minimal assumption of coherent evolution of modes after $a_{eq}$. Considering the uncertainty in the determination of initial spectra, $A_S^2$, from the CMB anisotropy, which is degenerate with the optical depth of re-ionization, we combine both $A_S^2$ and $g_X$ (where $X$ denotes each component of $\Phi$, $b$ and $\Theta_m$) with proper scaling for convenience as $G_X=g_XA_S/A_S^{*}$. Our result on measuring the coherent motion is independent of our choice of an arbitrary constant $A_S^{*\,2}$. The normalized density and coherent motion growth functions are given by $G_b$ and $G_{\Theta}$.

The transverse and radial distance components are given by $D_A$ and $H^{-1}$. Each galaxy is located using angular coordinates and redshift in galaxy redshift surveys. Because the clustering function of $\xi_s(\sigma,\pi)$ is measured in comoving distances, the fiducial cosmology should be applied to generate maps in comoving space. The transformation depends on the transverse and radial distances. Instead of recreating maps in comoving space, we obtain the approximate fiducial maps by rescaling the transverse and radial distances. This Alcock--Paczynski effect is described by the relation between ($D_A^{\rm fid},H^{-1\,\rm fid}$) and ($D_A^{\rm true},H^{-1\,\rm true}$), where 'fid' and 'true' denote the fiducial and fitting distances, respectively. The distance measures $\sigma^{\rm true}$ and $\pi^{\rm true}$ are transformed into the fiducial space as~\citep{Ballinger:1996cd},
\ba
\sigma^{\rm fid}&=&\frac{D_A^{\rm fid}}{D_A^{\rm true}}\sigma^{\rm true}, \nn \\
\pi^{\rm fid}&=&\frac{H^{-1\,\rm fid}}{H^{-1\,\rm true}}\pi^{\rm true} \,.
\ea
Accordingly,
\ba
s^{\rm fid} &=& \sqrt{\left(\frac{D_A^{\rm fid}}{D_A^{\rm true}}\right)^2\sigma^{2\,\rm true}+\left(\frac{H^{-1\,\rm fid}}{H^{-1\,\rm true}}\right)^2\pi^{2\,\rm true} }, \nn \\
\mu^{\rm fid} &=& \left(\frac{H^{-1\,\rm fid}}{H^{-1\,\rm true}}\pi^{\rm true}\right)\\
&&\Big{/} \sqrt{\left(\frac{D_A^{\rm fid}}{D_A^{\rm true}}\right)^2\sigma^{2\,\rm true}+\left(\frac{H^{-1\,\rm fid}}{H^{-1\,\rm true}}\right)^2\pi^{2\,\rm true} }\nn\,.
\ea
The measured $\xi^{\rm fid}$ is given in the fiducial parameter space with ($D_A^{\rm fid},H^{-1\,\rm fid}$), but the correlation function is calculated in the fitting parameters space with ($D_A^{\rm true},H^{-1\,\rm true}$). The theoretical $\xi$ with arbitrary ($D_A,H^{-1}$) is fitted to $\xi^{\rm fid}$ using rescaling in the above equations.

The spectra of the density and the velocity fields are naturally expected to receive nonlinear corrections. One of those corrections come from the random motion of galaxies, which results in the damping effect of the power spectrum amplitude. The Gaussian type of FoG function in Eq.~\ref{eq:Ggauss} is applied in this work. While the shape of FoG effect is given, the one-point PDF of the velocity $\sigma_p$ is set to be free parameter. We do not understand fully these non--perturbative damping effects at this time. Although we are not able to describe redshift distortions at much smaller scales, we vary $\sigma_p$ to fit data sets in which the leading order contribution of FoG is dominant. 

In summary, we have $G_b$ and $G_{\Theta}$ to describe growth functions, $D_A$ and $H^{-1}$ to fit distance measures, and $\sigma_p$ to model FoG effect. The shape of linear spectra and the form of FoG are assumed to be given by CMB experiments.

\subsection{The broadband Alcock--Paczynski test}

\begin{figure*}
\begin{center}
\resizebox{3.2in}{!}{\includegraphics{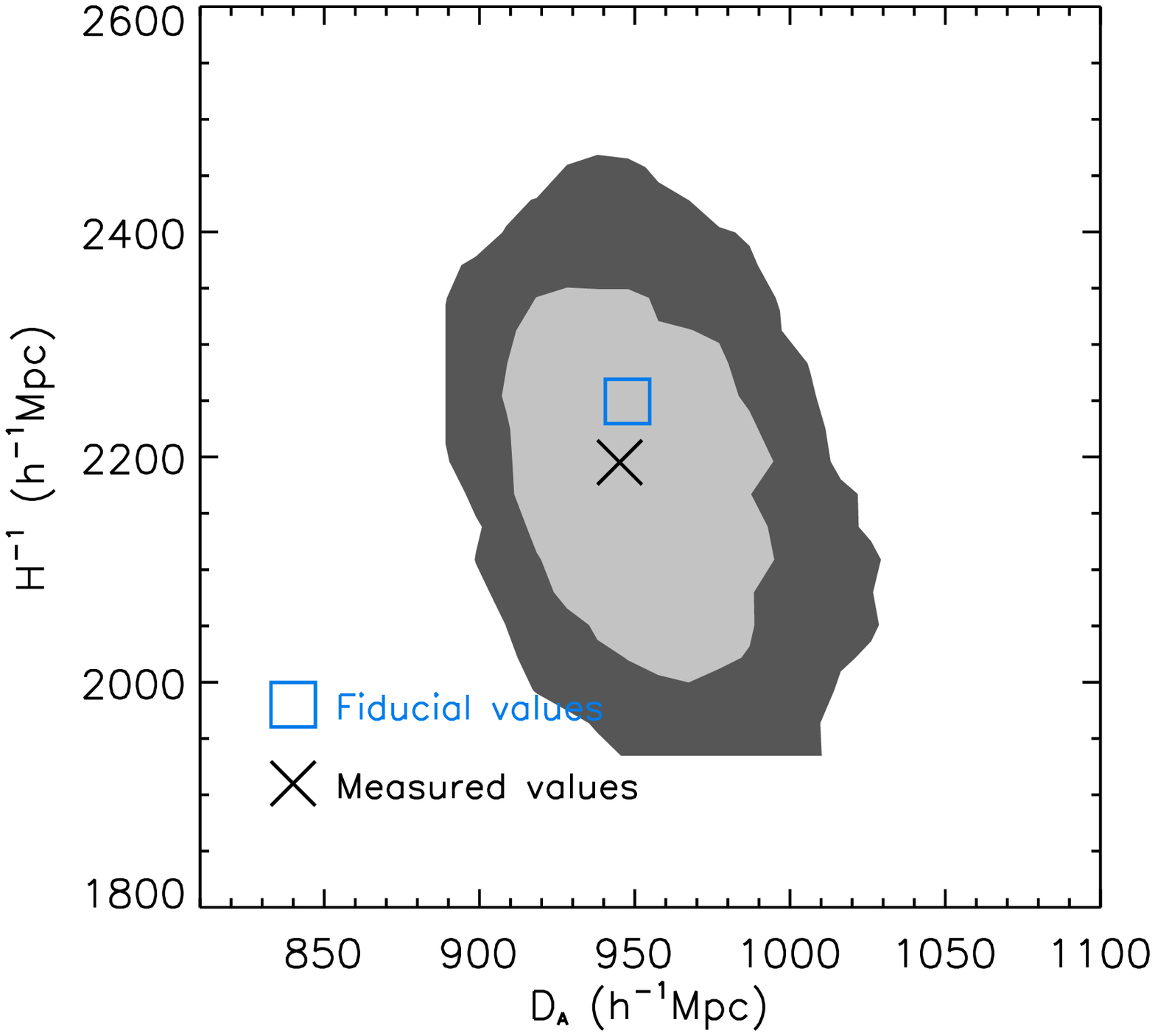}}\hfill
\resizebox{3.2in}{!}{\includegraphics{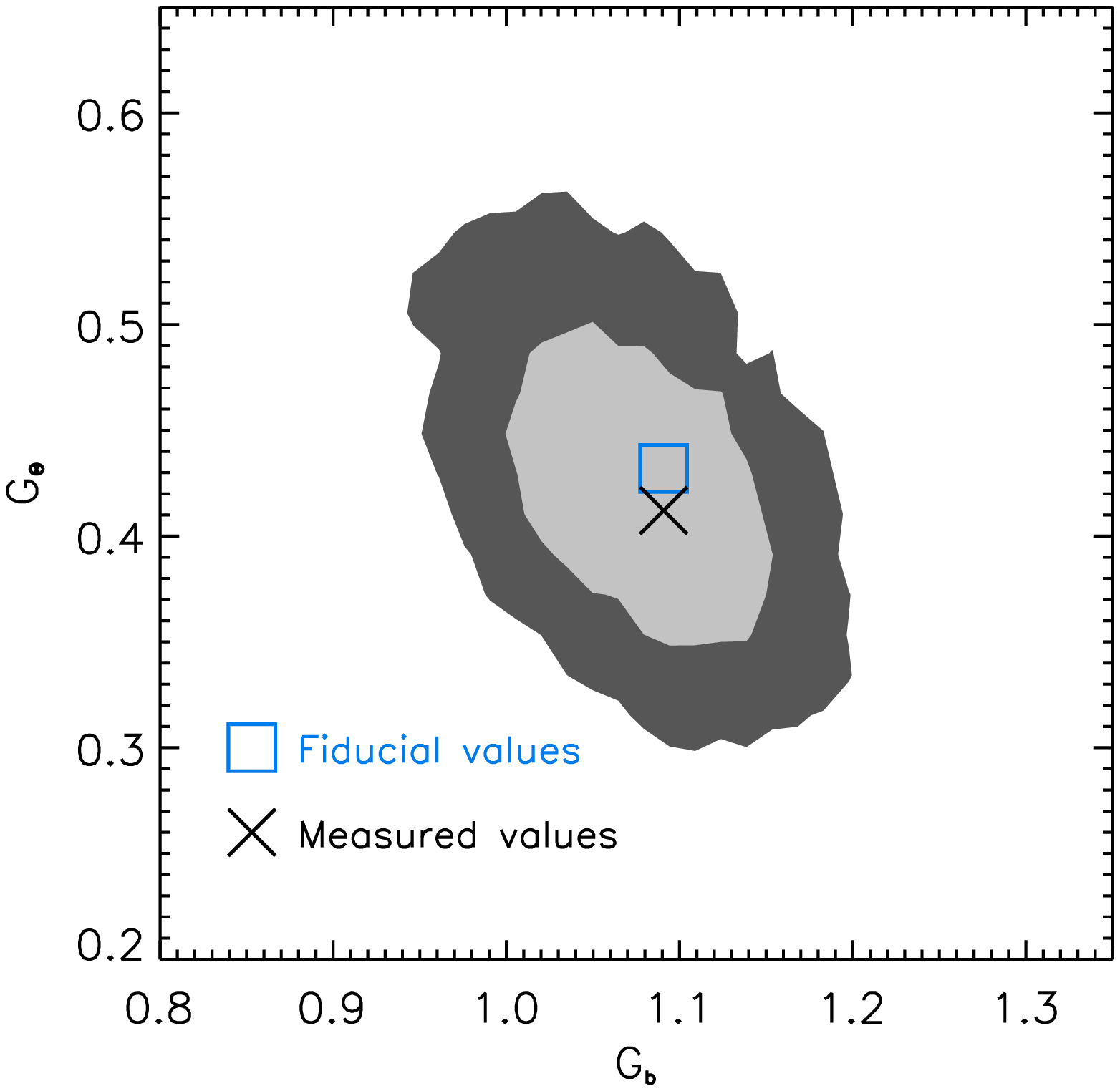}}
\end{center}
\caption{{\it (Left panel)} The measured $D_A$ and $H^{-1}$ are presented. The inner and outer contours represent $1-\sigma$ and $2-\sigma$ confidence levels respectively. The blue square point represents the fiducial values of $D_A$ and $H^{-1}$, and the black crossed point represents the measured value. {\it (Right panel)} The measured $G_b$ and $G_{\Theta}$ are presented. The inner and outer contours represent $1-\sigma$ and $2-\sigma$ confidence levels respectively. The blue square point represents the fiducial value of $G_{\Theta}$, and the black crossed point represents the measured value. }
\label{fig:confinal}
\end{figure*}

We present $\xi_s(\sigma,\pi)$ contours in the left panel of Fig.~\ref{fig:con_growth}. Three inner curves labeled by SI, SII and SIII represent $\xi_s(\sigma,\pi)$ contours without BAO peaks. The levels of contours are $\xi_s(\sigma,\pi)=0.2$, 0.06 and 0.16. Three outer curves labeled by BI, BII and BIII represent $\xi_s(\sigma,\pi)$ contours with BAO peaks. The levels of contours are $\xi_s(\sigma,\pi)=0.005$, 0.002 and $-0.001$. We connect all tip points of B--type contours to define the 2D BAO circle which is presented as the thick black solid curve in the Figure. In this subsection, the variation of the 2D BAO circle is presented in detail. 

The growth function $G_b$ is a dominant parameter of $\xi_0(s)$. It was reported in details the behavior of $\xi_s(\sigma,\pi)$ to the variation of $G_b$~\citep{Song:2010,Song:2011}. The variation of $G_b$ results in monopole amplification of S--type contours. Here, we pay attention to the variation of BAO peaks of B--type contours. Thin dash blue curves in the left panel of Fig.~\ref{fig:con_growth} represent $\xi_s(\sigma,\pi)$ contours with the increment of $G_b$. When the broadband shape of spectra is given, the observed BAO tip points mitigate on the same 2D BAO circle. Those are shifted counter--clockwise, while the 2D BAO circle remains unchanged. When $G_b$ decreases, the BAO tip points move clockwise. Again the 2D BAO circle itself is unchanged. 

The anisotropic amplification of S--type contours is generated with the increment of $G_{\Theta}$, because the cross--correlation of density and velocity is a leading term of $\xi_2(s)$~\citep{Song:2010,Song:2011}. For B--type contours, we observed the invariant 2D BAO circle as well to the variation of $G_{\Theta}$. The BAO tip points move toward the pivot point. The modified B--type contours are presented as thin blue dash contours in the right panel of Fig.~\ref{fig:con_growth}. The modified 2D BAO circle is presented as thick blue dash curve which is identical to the original circle. When $G_{\Theta}$ decreases, the BAO tip points run away from the pivot point on the same circle. 

If the correct distances are known, the tip points of BAO peaks form a invariant great circle regardless of galaxy bias and coherent motion.

When $D_A$ is unknown, the 2D BAO circle becomes uncertain transversely. We decrease $D_A$ by 10$\%$. The modified $\xi_s(\sigma,\pi)$ is presented with thin blue dash contours in the left panel of Fig.~\ref{fig:con_distance}. The 2D BAO circle is squeezed along the transverse direction. With the increment of $D_A$, the circle is stretched along the same direction. When the radial distance varies, the similar behaviors are observed along the $\pi$ direction. The blue thin dash curves in the right panel of Fig.~\ref{fig:con_distance} represent the modified $\xi_s(\sigma,\pi)$ when $H^{-1}$ decreases by 10$\%$. We connect the modified BAO tip points, and find a squeezed circle radially, represented by thick blue dash curve. Again, the radially stretched circle is expected with the increment of $H^{-1}$. 

Ratio between the observed transverse and radial distances varies with the assumed theoretical models. If the shape of an object is priorly known, then the ratio of the intrinsic transverse and radial distances gives a relation of the observed distance measures of $D_A$ and $H^{-1}$. It is a spirit of Alcock--Paczynski test~\citep{Alcock:1979mp}. We show that the 2D BAO circle provides us the object to be invariant to the change of growth functions of density and velocity fields, when the broadband shape of spectra is known. In addition, we show the orthogonal variation of 2D BAO circle to change of $D_A$ and $H^{-1}$, transversely and radially with respect to each other. Therefore, we dub this method as a broadband Alcock--Paczynski test. It is fortunate for us that those BAO peaks are formed at large scales about $s\sim 100\mpcoh$ in which non--perturbative effect on redshift distortions are well described with our theoretical model. Thus, in contrast to the classic method with the broadband but featureless power spectrum or correlation function \citep[e.g.,][]{Ballinger:1996cd,Matsubara:1996nf,Magira:1999bn}, the analysis taking account of BAO is expected to be much more powerful.

\section{Results}

\begin{table}
\begin{center}
\begin{tabular}{lccc}
Parameters & Fiducial values & Measurements \\
\hline
\hline
$D_A\,(\mpcoh)$       & $947.7$ & $945.1^{+33.0}_{-25.1}$  \\
\hline
$H^{-1}\,(\mpcoh)$   & $2249.4$ & $2195.3^{+98.8}_{-130.8}$  \\
\hline
$G_b$   & $-$ & $1.09^{+0.04}_{-0.07}$ \\
\hline
$G_{\Theta}$ &$0.43$ & $0.41^{+0.06}_{-0.05}$ \\
\hline
$\sigma_p\,(\mpcoh)$ &$-$ & $5.2^{+2.9}_{-3.9}$ \\
\hline
\end{tabular}
\end{center}
\caption{We present the fiducial and the measured values of $(G_b,G_{\Theta},D_A,H^{-1},\sigma_p)$ with the $1-\sigma$ confidence level errors of each realisation. The fiducial values of phenomenological parameters of $G_b$ and $\sigma_p$ are not given here.}
\label{tab:measurements}
\end{table}

In this section, we present the results of broadband Alcock--Paczynski test explained in the previous sections. We present measured values in Table~\ref{tab:measurements}, and discuss details below. The quoted errors on the best fit parameters are one sigma, after marginalizing over all other parameters. Throughout the analysis, we adopt the cut-off scales, $s=50\,h^{-1}$\,Mpc and $\pi=20\,h^{-1}$\,Mpc conservatively. The upper fitting range is $\sigma=150\mpcoh$ and $\pi=150\mpcoh$.

The angular diameter distance is measured to be $945.1^{+33.0}_{-25.1}$. The fiducial value of $D_A=947.7\mpcoh$ is within $1-\sigma$ confidence level of this measured value. The transverse distance is more relevant to the measured $\xi_s(\sigma,\pi)$ at $\pi\rightarrow 0$ bins where the contamination due to non--perturbative effect is minimized. If $D_A$ were not measured in precision, it would be caused more by the violation of broadband assumption of galaxy density spectra. The $D_A$ is degenerate more with $G_b$ than any other observable. The broadband assumption on galaxy density spectra is available when the galaxy bias remains scale--independent above cut--off scales. The precise measurement of $D_A$ in Table~\ref{tab:measurements} indicates that our assumption is correct.

While the transverse distance can be probed by another method, the radial distance is a unique outcome of Alcock--Paczynski test. The radial distance is measured to be $H^{-1}=2195.3^{+98.8}_{-130.8}\mpcoh$, when the fiducial $H^{-1}$ is $2249.4\mpcoh$. The fiducial value of $H^{-1}$ is also well--within $1-\sigma$ confidence level. The measurement of $H^{-1}$ is influenced more by radial bins at $\mu\rightarrow 1$ where the contamination due to non--perturbative effect is maximized. Therefore, the higher systematic uncertainty is expected to plague through the measured $H^{-1}$. Fortunately, it is not severe in our case using BOSS--like simulation. Although $H^{-1}$ measurement can be biased due to this systematic uncertainty, the effect is buried under the much bigger statistical error. We provide the contour plot of $D_A$ and $H^{-1}$ in the left panel of Fig.~\ref{fig:confinal}. The fiducial values are represented by a blue square point, the measured values are represented by a black cross point. The inner and outer contours represent the $1-\sigma$ and $2-\sigma$ confidence levels respectively. The measured values of $D_A$ and $H^{-1}$ are located well within $1-\sigma$ bound. We add a comment on the relation between the measured distances and the sound horizon at last scattering surface, $r_s$. Strictly speaking, distances are measured in $D_Ar_s^{\rm fid}/r_s$ and $H^{-1}r_s^{\rm fid}/r_s$. But, with the given CMB prior, $r_s$ is determined by CMB prior in higher accuracy than fractional errors for measured $D_A$ by RSD.

The accurate measurement of coherent motion is very important to test modified gravity model cosmologically. The model independent test of general relativity is only available by probing both geometrical perturbations and matter fluctuations separately. While the geometrical perturbations are measured by cosmic shear experiments, the unbiased probe of matter fluctuations is only possible by measuring coherent motions using redshift distortions. It is known that the measurement of coherent motions is difficult, because of non--linear physics contamination. The improved theoretical models of redshift distortions is reported to be an appropriate resolution to overcome this problem for the case of dark matter particles~\citep{Song:2013}. Here, we show that the coherent growth function $G_{\Theta}$ is probed precisely for the case of CMASS samples as well. The $G_{\Theta}$ is measured to be $0.41^{+0.06}_{-0.05}$, when the fiducial value is $G_{\Theta}=0.43$. We present it graphically in the right panel of Fig.~\ref{fig:confinal}, in order to highlight the precision level. The measured $G_{\Theta}$ can be converted into structure formation parameter $f\sigma_8$~\citep{Song:2008qt} as $f\sigma_8=0.42^{+0.07}_{-0.06}$. The fiducial $f\sigma_8$ is 0.44. The fractional error to measure $G_{\Theta}$ is about $10\%$. The constraint on coherent motion is poorer by high degeneracy with FoG effect at linear regime. More precise measurement of $G_{\Theta}$ is only possible by better understanding of FoG effect.

Unlike observables explained above, there is no fiducial $G_b$ known. The square point in the right panel of Fig.~\ref{fig:confinal} is presented with the measured $G_b=1.09^{+0.04}_{-0.07}$. The growth function $G_{\delta_m}$ of dark matter is 0.56 at $z=0.57$. If the galaxy bias is constant, then the measured galaxy bias can be estimated as $b=1.91^{+0.08}_{-0.14}$. The measured galaxy biases are reported in many literature for the same simulation. It ranges from 1.8 to 2, depending on the assumption made in each method. Our galaxy bias is estimated from the decomposed galaxy density spectra which are supposed to be close to spectra in real space. Therefore, it should agree with the value which is measured in real space of the simulation. This value is about $b=1.9$ published in~\cite{Manera:2013}. Our measurement agree with it. 

Finally, we comment on the FoG parameter $\sigma_p$. The measured result is $5.2^{+2.9}_{-3.9}\mpcoh$, which is higher than the linear theory estimate of one--dimensional velocity dispersion, $\sigma_p=3\mpcoh$. The measured velocity dispersion is a factor of 1.5 higher than the linear prediction. The larger value of $\sigma_p$ might be caused by the higher-order nonlinear correction. Another possibility is the functional form of FoG, which is indeed not precisely known. The discrepancy from Gaussian FoG function may give us impacts on the measured $\sigma_p$. If more precise expression on FoG effect is available, then tighter constraints on observables will be made. There are many efforts to formulate non--perturbative effect~\citep{Reid:2011ar,Okumura:2011pb}.

\section{Conclusion}

Redshift distortions provide a large cosmological object which is appropriate for cosmological tool to measure geometric distances via the Alcock--Paczynski effect. The tip points of BAO peaks on $\xi_s(\sigma,\pi)$ clustering function form a great circle called as the 2D BAO circle. When the shape of spectra is pre--determined by CMB experiments, the 2D BAO circle exhibits characteristic anisotropies which remain unchanged to the variations of the unknown galaxy bias and the coherent motion growth function. This feature enables us to probe both transverse and radial distances uniquely. The 2D BAO circle is observed to be altered transversely to the variation of $D_A$, and radially to the variation of $H^{-1}$. It becomes a reliable large object to test Alcock--Paczynski effect cosmologically. We dub it as a broadband Alcock--Paczynski test in this paper.

The success of this test depends on theoretical models of redshift distortions. We show that BAO peaks of $\xi_s(\sigma,\pi)$ are precisely reproduced by the improved redshift distortion model at linear regime. This theoretical model is limited by unknown non--perturbative effect. We ignore the measured $\xi_s(\sigma,\pi)$ at small bins in which the contamination caused by non--perturbative effect exceeds more than the first order approximation of Gaussian FoG function. Our cut--off scales are $s=50\mpcoh$ and $\pi=20\mpcoh$ which are enough to include most BAO features, while non--perturbative contamination is safely removed. We reproduce successfully all fiducial values of $(D_A,H^{-1},G_{\Theta})$ well within $1-\sigma$ confidence level. Our results support redshift distortions as a trustable tool to measure key observables to test the dark energy and modify gravity models, even with the conservative bound at linear regime. We note that those constraints will be tightened with better understanding of FoG function in the future.

\section*{Acknowledgments}

Numerical calculations were performed by using a high performance computing cluster in the Korea Astronomy and Space Science Institute. We would like to thank Marc Manera for providing us with his mock catalogs before publication. This research was partly supported by WCU grant R32-10130 and Ewha Womans University research fund 1-2008-2935-001-2 (T.O.) and a Grant-in-Aid for Scientific Research from the Japan Society for the Promotion of Science (A.T, No.~24540257).

\appendix 

\section{singular value decomposition method}

We explain details of the SVD method which is used in this work. When using SVD the $\chi^2$ value becomes more difficult to interpret as it changes as one cuts the noisiest eigenvalues.  However we establish that the reduced $\chi^2$ converges to a constant value above 250 modes. To be conservative we use 345 out of 400 available modes.

\begin{figure}
\begin{center}
\resizebox{3.2in}{!}{\includegraphics{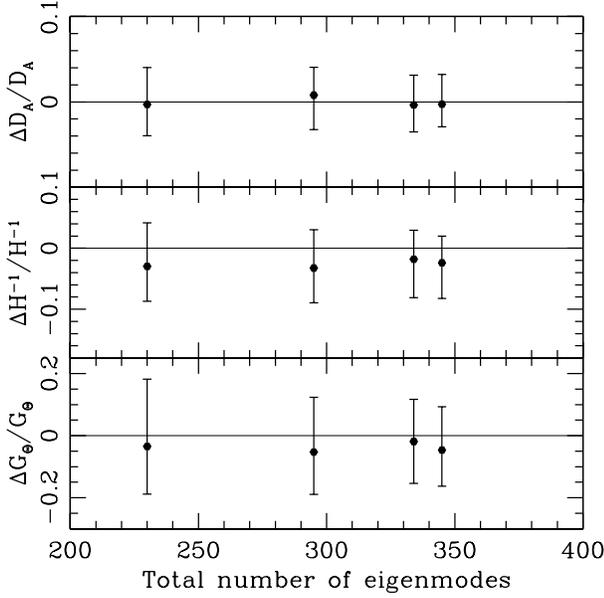}}
\end{center}
\caption{The fractional errors of $D_A$, $H^{-1}$ and $G_{\Theta}$ are presented using different eigenmode cut--off from the top to bottom panels.}
\label{fig:SVDtest}
\end{figure}

We discuss the caveat when we use SVD method to calculate covariance matrix. As $\chi^2$ changes as one cuts the noisiest eigenvalues, the convergence should be tested when the eignemode cut is applied. We refit the $\xi_s(\sigma,\pi)$ measured data using different eignemode cut. The total number of eigen modes varies from 250 to 350. The measured values and the fractional errors of $D_A$, $H^{-1}$ and $G_{\Theta}$ are presented in~Fig.\ref{fig:SVDtest}. The measured values are consistent within the error budget, and the fraction errors approximately converges at the total number of modes of 345.

%\bibliographystyle{mn2e} 
%\bibliography{bibYSS}

%\begin{thebibliography}{53}

\end{document}